\documentclass[a4paper]{papertex}
\pdfoutput=1
\usepackage{etex}

\usepackage[utf8]{inputenc}

\usepackage[table]{xcolor}
\usepackage{booktabs} 
\usepackage{graphicx}
\usepackage{numprint}
\usepackage{wrapfig}
\usepackage{tabularx}
\usepackage{mdframed}
\usepackage{xspace}
\usepackage{cellspace}

\newcommand{\TODO}[1]{\textcolor{red}{todo: #1}}\newcommand\todo\TODO
\newcommand{\TODOSU}[1]{\textcolor{green}{todo: SU: #1}}\newcommand\todosu\TODOSU
\usepackage{listings}

\newcommand\nbpatches{\numprint{15}\xspace}
\newcommand\nbpatchesmerged{\numprint{5}\xspace}
\newcommand{\ra}[1]{\renewcommand{\arraystretch}{#1}}
\lstdefinelanguage{diff}{
  language=java,
  basicstyle=\ttfamily\scriptsize,
  sensitive=true,
  numbers=none,
  morecomment=[f][\color{gray}][0]{diff},
  morecomment=[f][\color{gray}][0]{index},
  morecomment=[f][\color{blue}][0]{@@},
  morecomment=[f][\color{magenta}][0]{***},
  morecomment=[f][\color{violet}][0]{!},
  morecomment=[f][\color{red!60!black}][0]{-},
  morecomment=[f][\color{green!60!black}][0]{+},
  morecomment=[f][\color{magenta}][0]{---},
  morecomment=[f][\color{magenta}][0]{+++},
  morecomment=[f][\color{gray}][0]{Binary},
  morecomment=[f][\color{gray}][0]{Only},
  morecomment=[f][\color{gray}][0]{old},
  morecomment=[f][\color{gray}][0]{new},
  morecomment=[f][\color{gray}][0]{rename},
  morecomment=[f][\color{gray}][0]{similarity},
  morecomment=[f][\color{gray}][0]{deleted},
  morecomment=[f][\color{magenta}][0]{***************},
  morecomment=[f][\color{red!60!black}][0]<,
  morecomment=[f][\color{green!60!black}][0]>,
  morecomment=[f][\color{blue}][0]{0},
  morecomment=[f][\color{blue}][0]{1},
  morecomment=[f][\color{blue}][0]{2},
  morecomment=[f][\color{blue}][0]{3},
  morecomment=[f][\color{blue}][0]{4},
  morecomment=[f][\color{blue}][0]{5},
  morecomment=[f][\color{blue}][0]{6},
  morecomment=[f][\color{blue}][0]{7},
  morecomment=[f][\color{blue}][0]{8},
  morecomment=[f][\color{blue}][0]{9},
}[comments]

\title{}

\author{Martin Monperrus et al.}

\begin{document}

\begin{news}{3}{Human-competitive Patches in Automatic Program Repair with Repairnator}{Martin Monperrus, Simon Urli, Thomas Durieux, Matias Martinez, Benoit Baudry, Lionel Seinturier}{Software engineering research at KTH Royal Institute of Technology \& Inria}{hlabel i}

\textbf{Repairnator is a bot. It constantly monitors software bugs discovered during continuous integration of open-source software and tries to fix them automatically. If it succeeds to synthesize a valid patch, Repairnator proposes the patch to the human developers, disguised under a fake human identity. To date, Repairnator has been able to produce \nbpatchesmerged patches that were accepted by the human developers and permanently merged in the code base. This is a milestone for human-competitiveness in software engineering research on automatic program repair.}

\vspace{.5cm}

Program repair research pursues the idea that algorithms can replace humans to fix software bugs \cite{Monperrus2015}.
A bug fix is a patch that inserts, deletes or modifies source code. For example, in the following patch, the developer has modified the condition of the if statement:

\begin{lstlisting}
- if (x < 10)
+ if (x <= 10)
  foo();
\end{lstlisting}

A program repair bot is an artificial agent that tries to synthesize source code patches. It analyzes bugs and produces patches, in the same way as human developers involved in software maintenance activities.
This idea of a program repair bot is disruptive, because today humans are responsible for fixing bugs.
In others words, we are talking about a bot meant to (partially) replace human developers for tedious tasks. 

When a bot tries to achieve a task usually done by humans, it is known as a human-competitive task \cite{koza2010human}. 
The empirical evaluations of program repair research \cite{defects4j-repair}  show that current program repair systems are able to synthesize patches for real bugs in real programs. 
However, all those patches were synthesized for past bugs, for bugs that were fixed by human developers in the past, usually years ago. While this indicates the technical feasibility of program repair, this fails to show that program repair is human-competitive.

\subsection*{Human-competitiveness}

To demonstrate that program repair is human-competitive, a program repair bot has to find a high-quality patch before a human does so. 
In this context, a patch can be considered to be human-competitive if it satisfies the two conditions of timeliness and quality. 
Timeliness refers to the fact that the system must find a patch before the human developer. In other words, the prototype system must produce patches in the order of magnitude of minutes, not days.  
Also, the patch generated by the bot must be correct-enough, of similar quality -- correct and readable -- compared to a patch written by a human.
Note that there are patches that look correct from the bot's point of view, yet that are incorrect (this is known as overfitting patches in the literature \cite{Smith2015,defects4j-repair}). Those patches are arguably not human-competitive, because humans would never accept them in their code base.

Consequently, for a patch to be human-competitive 
1)  the bot has to synthesize the patch faster than the human developer
2) the patch has to be judged good-enough by the human developer and permanently merged in the code base. 

There is one more aspect to consider. It has been shown that human engineers do not accept contributions from bots as easily as contributions from other humans, even if they are strictly identical \cite{murgia2016among}. The reason is that humans tend to have a priori biases against machines, and are more tolerant to errors if the contribution comes from a human peer. In the context of program repair, this means that developers may put the bar higher on the quality of the patch, if they know that the patch comes from a bot. This would impede our quest for a human-competitiveness proof in the context of program repair.

\begin{table*}
\begin{minipage}{1\textwidth}
\ra{1.3}
\rowcolors{2}{gray!20}{white}
\begin{tabularx}{\textwidth}{p{21mm}  X  p{85mm}}\toprule
\textbf{Date} & \textbf{Contribution}  & \textbf{Developer comment}  \\ \hline

Jan 12, 2018 & \href{https://github.com/aaime/geowebcache/pull/1}{aaime/geowebcache/pull/1} & ``Thanks for the patch!'' \\

Mar 23, 2018 & \href{https://github.com/parkito/BasicDataStructuresAndAlgorithms/pull/3}{parkito/BasicDataCodeU[...]/pull/3} & ``merged commit 140a3e3 into parkito:develop'' \\

April 5, 2018 & \href{https://github.com/dkarv/jdcallgraph/pull/2}{dkarv/jdcallgraph/pull/2} & ``Thanks!''  \\

May 3, 2018 & \href{https://github.com/eclipse/ditto/pull/151}{eclipse/ditto/pull/151} & ``Cool, thanks for going through the Eclipse process and for the fix.''  \\

June 25, 2018 & \href{https://github.com/donnelldebnam/CodeU-Spring-2018-29/pull/59}{donnelldebnam/CodeU[...]/pull/151} & ``Thanks!!''\\

\bottomrule
\end{tabularx}
\end{minipage}
\caption{Human-competitive contributions on Github: patches synthesized by the Repairnator robot and accepted by the human developer.}
\label{tab:repainator-accepted-patches}
\end{table*}

To overcome this problem, we have decided early in the project that all Repairnator patches would be proposed under a fake human identity. We have created a GitHub user, called Luc Esape, who is presented as software engineer in our research lab. Luc has a profile picture and looks like a junior developer, eager to make open-source contributions on GitHub. Now imagine Repairnator, disguised as Luc Esape proposing a patch: the developer reviewing it
thinks that she is reviewing a human contribution.
This camouflage is required to test our scientific hypothesis of human-competitiveness. Now, for sake of ethics, the real identity of Luc has been disclosed on each of his pull-requests.

\subsection*{Automatic Repair and Continuous Integration}

Continuous integration, aka CI, is the idea that a server compiles the code and runs all tests for each commit made in the version control system of a software project (e.g. Git).
In CI parlance, there is a \emph{build} for each commit.
A build contains the information about the source code snapshot used (e.g. a reference to a Git commit), the result of compilation and test execution (e.g. fail or success), and an execution trace log. 
A build is said to be failing if compilation fails or at least one test case fails. It has been shown that approximately one out of 4 builds fails, and that the most common cause for build failure is a test failure \cite{vassallo2017tale}.

The key idea of Repairnator is to automatically generate patches that repair build failures, then to show them to human developers, to finally see whether those human developers would accept them as valid contributions to the code base. If this happens, that would be evidence of human-competitiveness in program repair.

This setup --automatically repairing build failures happening in continuous integration -- is particularly appropriate and timely for the following reasons.
First, build failures satisfy the core problem statement of test-suite based program repair~\cite{Monperrus2015}, where bugs are specified as a failing test-cases, and those failing test cases are used to drive the automated synthesis of a patch~\cite{Monperrus2015}.
Second, it allows comparing the bots and humans on a fair basis: when a failing test is discovered on the continuous integration server, the human developer and the bot are informed about it, \emph{at the exact same time}. This test failure notification is the starting point of the human vs. bot competition.

Repairnator's focus on build failures is unique, but it fits in the big picture of intelligent bots for software \cite{Lebeuf2017}. For instance, Facebook has a tool called SapFix that repairs bugs found with automated testing. Also related,  the DARPA Cyber Grand Challenge bot attackers and defenders try to be human-competitive with respect to  security experts. 

\subsection*{Repairnator in a Nutshell}

In 2017-2018, we have designed, implemented and operated
Repairnator, a bot for automated program repair.
Repairnator is specialized to repair build failures happening during continuous integration.
It constantly monitors thousands of commits being pushed to the GitHub code hosting platform, and analyzes their corresponding builds. Every minute, it launches new repair attempts in order to fix bugs before the human developer.
It is designed to go as fast as possible because it participates to a race: if Repairnator finds a patch before the human developer, it is human-competitive.

Let us now give an overview of how the Repairnator bot works.

\noindent\includegraphics[width=\columnwidth]{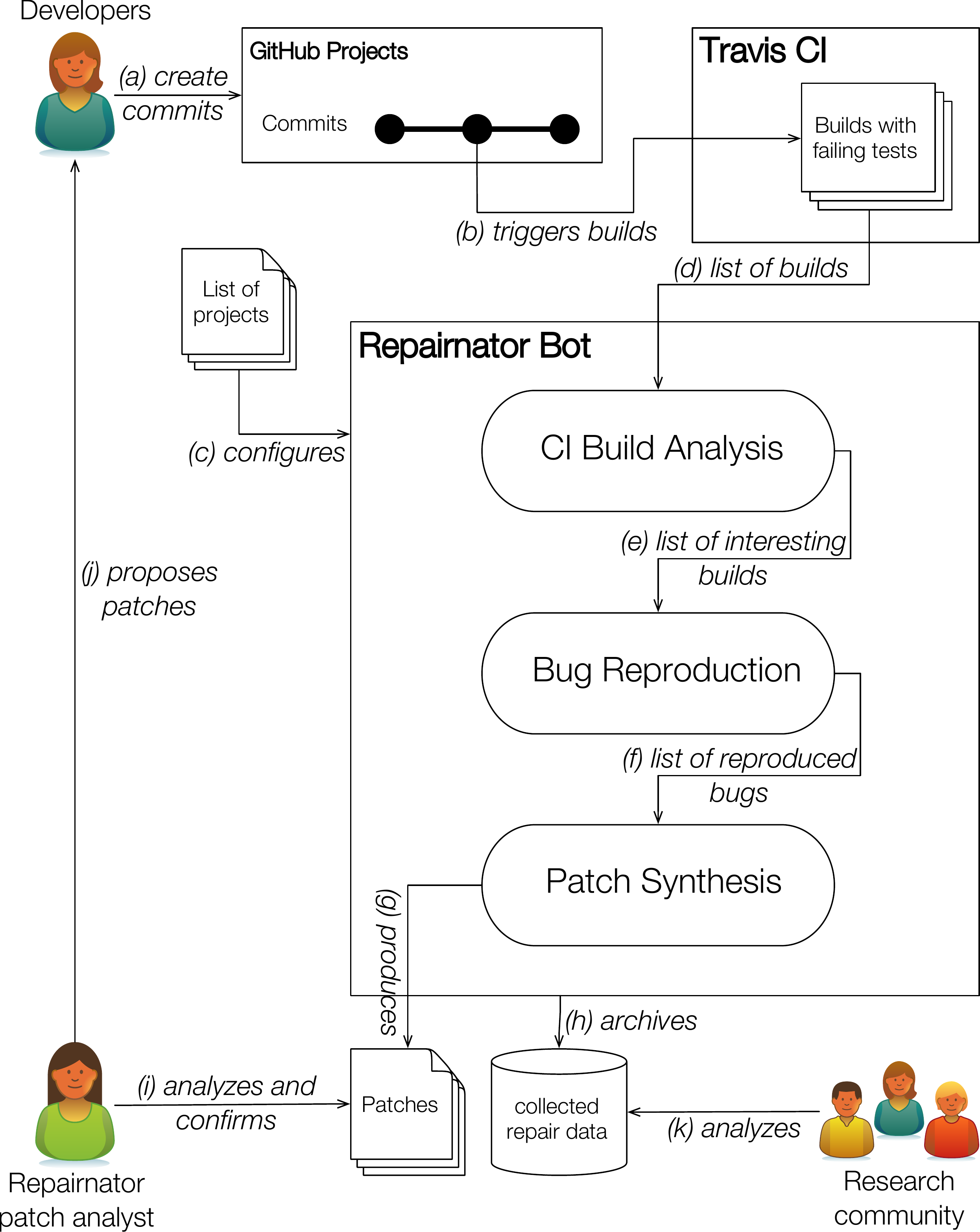}

The primary input of Repairnator are continuous integration builds, triggered by commits made by developers (top part of the figure, \emph{(a)} and \emph{(b)}) based on GitHub projects \emph{(a)}.
The outputs of Repairnator are two-fold:
\emph{(1)} it automatically produces patches for repairing failing builds \emph{(g)}, if any;
\emph{(2)} it collects valuable data for future program repair research \emph{(h)} and  \emph{(k)}.
Permanently, Repairnator monitors all continuous activity from GitHub projects \emph{(c)}.
The CI builds are given as input to a three stage pipeline: 
\emph{(1)} a first stage collects and analyzes CI build logs \emph{(e)};
\emph{(2)} a second stage aims at locally reproducing the build failures that have happened on CI \emph{(f)}; 
\emph{(3)} a third stage runs different program repair prototypes coming from the latest academic research. 
When a patch is found, a Repairnator project member performs a quick sanity check, in order to avoid wasting valuable time of open-source developers.
\emph{(i)} If she finds the patch non-degenerated, she then proposes it to the original developers of the project as a pull-request on GitHub \emph{(j)}. 
The developers then follow their usual process of code-review and merge.

Repairnator has to operate in a given software ecosystem. Due to our expertise with Java in past research projects, the prototype implementation of Repairnator focuses on repairing software written in the Java programming language, built with the Maven toolchain, in open-source projects hosted on GitHub, which use the Travis CI continuous integration platform.

\subsection*{Expedition Achievements}
We have been operating Repairnator since January 2017, in three different phases.
During one month, in January 2017, we performed a pilot experiment with a initial version of the prototype.
From February 1st, 2017 to December 31th, 2017, we ran Repairnator with a fixed list of \numprint{14,188} projects, we call it ``Expedition \#1''.
From January 1st 2018 to June 31th 2018, Repairnator has monitored the Travis CI build stream in real time, we call it ``Expedition \#2''

The main goal of the pilot experiment was to validate our design and initial implementation. We found that our prototype is capable of performing approximately 30 repair attempts per day, given our computing resources. 
More importantly, this pilot experiment validated our core technological assumptions:
a significant proportion of popular open-source projects use Travis and the majority of them use Maven as build technology. This meant we would have a fair chance of reaching our goal of synthesizing a human-competitive patch in that context.

During Expedition \#1, whose results are presented in details in~\cite{urli:hal-01691496}, Repairnator has analyzed \numprint{11,523} builds with test failures.
For \numprint{3,551} of them (30.82\%), Repairnator was able to locally reproduce the test failure.
Out of \numprint{3,551} repair attempts, Repairnator found \nbpatches patches that could make the CI build pass. However, our patch analysis revealed that none of those patches were human-competitive because they came too late (Repairnator produced a patch after the human developer) or were of low quality (they made the build successful coincidentally).

Expedition \#2 is the successful one. It has shown that program repair technology has crossed the frontier of human-competitiveness. Repairnator has produced 5 patches that meet the criteria of human-competitiveness defined above:
1) the patches were produced before the human ones, 2) a human developer accepted the patches as valid contributions, and the patches were merged in the main code base.
The information about those five milestone patches are shown in \autoref{tab:repainator-accepted-patches} and we now discuss the first of them.

The first patch merged by our program repair bot was accepted by a human developer on Jan 12th, 2018.
Here is the story:
on Jan 12th 2018 at 12:28pm, a build was triggered on project aaime/geowebcache\footnote{\url{https://travis-ci.org/GeoWebCache/geowebcache/builds/328076497}}. The build failed after 2 minutes of execution, because two test cases were in error.
Fourty minutes later, on Jan 12th 2018 at 1:08pm, Repairnator detected the failing build during its regular monitoring, and started to run the available program repair systems configured in Repairnator.
Ten minutes later, at 1:18pm, it found a patch. 

On Jan 12th 2018, at 1:35pm, a Repairnator team member  took the patch generated by Repairnator, and validated the opening of the corresponding pull-request on GitHub. On Jan 12th 2018, at 2:10pm, the developer accepted the patch, and merged it with a comment:  “Weird, I thought I already fixed this... maybe I did in some other place. Thanks for the patch!”.
That was the first patch produced by Repairnator and accepted as a valid contribution  by a human developer, definitively merged in the code base.
In other words, Repairnator was human-competitive for the first time.

After 6 more months of operation, Repairnator has had 5 patches merged by human developers, which are all listed in \autoref{tab:repainator-accepted-patches}.

Overall, the Repairnator project has fullfilled its mission. It has shown that program repair can be considered as human-competitive: Repairnator has found patches 1) before the humans, 2) that were considered of good quality by humans themselves.

\subsection*{The Future}
In addition to showing that program repair is human competitive, the Repairnator project has provided a wealth of information about bugs and continuous integration, and about the current shortcomings of program repair research, presented in \cite{urli:hal-01691496}. 

Let us dwell on one point in particular, the question of intellectual property. On May 3rd, 2018, Repairnator produced a good patch for GitHub project eclipse/ditto. Shortly after having proposed the patch, one of the developers asked \emph{``We can only accept pull-requests which come from users who signed the Eclipse Foundation Contributor License Agreement.''}. We were puzzled because a bot cannot physically or morally sign a license agreement and is probably not entitled to do so. Who owns the intellectual property and responsibility of a bot contribution: the robot operator, the bot implementer or the repair algorithm designer? This is one of the interesting questions uncovered by the Repairnator project.

We believe that Repairnator prefigures a certain future of software development, where bots and humans will smoothly collaborate and even cooperate on software artifacts.

\bibliographystyle{abbrv}
\bibliography{bibliography} 

\begin{thebibliography}{1}

\bibitem{koza2010human}
J.~R. Koza.
\newblock {Human-competitive Results Produced by Genetic Programming}.
\newblock {\em Genetic Programming and Evolvable Machines}, 11(3-4):251--284,
  2010.

\bibitem{Lebeuf2017}
C.~Lebeuf, M.~D. Storey, and A.~Zagalsky.
\newblock Software bots.
\newblock {\em {IEEE} Software}, 35:18--23, 2018.

\bibitem{defects4j-repair}
M.~Martinez, T.~Durieux, R.~Sommerard, J.~Xuan, and M.~Monperrus.
\newblock {Automatic Repair of Real Bugs in Java: a Large-scale Experiment on
  the Defects4j Dataset}.
\newblock {\em Empirical Software Engineering}, pages 1--29, 2016.

\bibitem{Monperrus2015}
M.~Monperrus.
\newblock {Automatic Software Repair : a Bibliography}.
\newblock {\em ACM Computing Surveys}, 2017.

\bibitem{murgia2016among}
A.~Murgia, D.~Janssens, S.~Demeyer, and B.~Vasilescu.
\newblock {Among the machines: Human-bot interaction on social q{\&}a
  websites}.
\newblock In {\em Proceedings of the 2016 CHI Conference Extended Abstracts on
  Human Factors in Computing Systems}, pages 1272--1279. ACM, 2016.

\bibitem{Smith2015}
E.~K. Smith, E.~T. Barr, C.~{Le Goues}, and Y.~Brun.
\newblock {Is the cure worse than the disease? overfitting in automated program
  repair}.
\newblock In {\em Proceedings of the 2015 10th Joint Meeting on Foundations of
  Software Engineering}, pages 532--543, 2015.

\bibitem{urli:hal-01691496}
S.~Urli, Z.~Yu, L.~Seinturier, and M.~Monperrus.
\newblock {How to Design a Program Repair Bot? Insights from the Repairnator
  Project}.
\newblock In {\em {ICSE 2018 - 40th International Conference on Software
  Engineering, Track Software Engineering in Practice}}, 2018.

\bibitem{vassallo2017tale}
C.~Vassallo, G.~Schermann, F.~Zampetti, D.~Romano, P.~Leitner, A.~Zaidman,
  M.~Di~Penta, and S.~Panichella.
\newblock {A Tale of {CI} Build Failures: An Open-source and a Financial
  Organization Perspective}.
\newblock In {\em International Conference on Software Maintenance and
  Evolution}, 2017.

\end{thebibliography}

\end{news}

\end{document}